\documentclass[sigconf, screen, nonacm]{acmart}

\usepackage{amsmath,amsfonts}
\usepackage{algorithmic}
\usepackage{graphicx}
\usepackage{textcomp}
\usepackage{xcolor}
\usepackage{tikz}
\usepackage{subcaption}
\usetikzlibrary{positioning,shapes.geometric,arrows.meta,calc,fit,backgrounds,decorations.pathmorphing}
\usepackage{orcidlink}
\usepackage[nolist]{acronym}
\usepackage[capitalize]{cleveref}
\usepackage{tikz}
\usepackage{xcolor}
\usepackage[inline]{enumitem}

\definecolor{naivered}{HTML}{E15759}
\definecolor{alignedblue}{HTML}{4E79A7}
\definecolor{readygreen}{HTML}{59A14F}
\definecolor{paneltint}{HTML}{F5F5F5}
\definecolor{cachegray}{HTML}{E8E8E8}

\Crefname{section}{Section}{Sections}
\crefname{section}{Sec.}{Secs.}
\Crefname{table}{Table}{Tables}
\crefname{table}{Tab.}{Tabs.}
\Crefname{figure}{Figure}{Figures}
\crefname{figure}{Fig.}{Figs.}
\crefname{lstlisting}{List.}{Lists.}

\definecolor{naivered}{HTML}{E15759}
\definecolor{alignedblue}{HTML}{4E79A7}
\definecolor{cachegray}{HTML}{D0D0D0}
\def\BibTeX{{\rm B\kern-.05em{\sc i\kern-.025em b}\kern-.08em
    T\kern-.1667em\lower.7ex\hbox{E}\kern-.125emX}}
\begin{acronym}
    \acro{MoQ}{Media over QUIC}
    \acro{MOQT}{Media Over QUIC Transport}
    \acro{ABR}{Adaptive Bitrate}
    \acro{HEVC}{High Efficiency Video Coding}
    \acro{CMAF}{Common Media Application Format}
    \acro{MSF}{MOQT Streaming Format}
    \acro{CMSF}{CMAF compliant MOQT Streaming Format}
    \acro{HAS}{HTTP Adaptive Streaming}
    \acro{DASH}{Dynamic Adaptive Streaming over HTTP}
    \acro{HLS}{HTTP Live Streaming}
    \acro{CDN}{Content Delivery Network}
    \acro{PTS}{Presentation Timestamp}
    \acro{MSE}{Media Source Extensions}
    \acro{GOP}{Group of Pictures}
    \acro{LL-DASH}{Low-latency DASH}
    \acro{HoL}{head-of-line}
    \acro{IETF}{Internet Engineering Task Force}
    \acro{TSA-SWITCH}{Time Shift-Aware SWITCH}
    \acro{LIC}{Learned Image Compression}
    \acro{INR}{Implicit Neural Representation}
    \acro{QP}{Quantization Parameter}
    \acro{MSE}{Mean Squared Error}
    \acro{SSIM}{Structural Similarity Index Measure}
    \acro{PSNR}{Peak Signal to Noise Ratio}
    \acro{VMAF}{Video Multi-Method Assessment\\Fusion}
    \acro{LPIPS}{Learned Perceptual Image Patch Similarity}
    \acro{UI-LIC}{Unified Interface for Learned Image Compression}
\end{acronym}

\def\name{\ac{UI-LIC}}

\AtBeginDocument{%
  \providecommand\BibTeX{{%
    Bib\TeX}}}

\begin{document}

\title{UI-LIC: A Unified Framework for Evaluating Learned Image Compression Models}

\author{Nicholas J. Nolen}
\email{Nicholas_Nolen1@baylor.edu}
\orcid{0009-0007-2640-7713}
\affiliation{%
  \institution{Baylor University}
  \city{Waco}
  \state{Texas}
  \country{USA}
}

\author{Luc Trudeau}
\email{luc.trudeau@gmail.com}
\orcid{0000-0002-4292-5875}
 \affiliation{%
   \institution{Université du Québec à Rimouski}
   \city{Rimouski}
   \state{Québec}
   \country{Canada}
 }

\author{Andrew C. Freeman}
\email{andrew_freeman@baylor.edu}
\orcid{0000-0002-7927-8245}
\affiliation{%
  \institution{Baylor University}
  \city{Waco}
  \state{Texas}
  \country{USA}
}

\begin{abstract}
 The evaluation and comparison of Learned Image Compression (LIC) systems is complicated by heterogeneous software stacks, varying training conditions, and divergent evaluation methodologies. To address these challenges, we introduce UI-LIC, an open-source software framework for evaluating LIC models. We integrate six high-performance LIC models, and provide a centralized controller for performing training, inference, and analysis with shared configuration parameters. Our GUI program offers a streamlined interface to evaluate these models  alongside traditional video intra-frame encoders, equalizing the compressed bitrates and calculating quality metrics such as PSNR, SSIM, VMAF, and LPIPS. Finally, we provide an interactive image analyzer with configurable quality heatmap overlays. Our framework lowers barriers to further LIC research, unlocking comparative metrics and subjective analysis with a single setup command. The open-source software is released under the MIT license and is available at \hyperlink{https://github.com/BaylorMultimediaLab/UI-LIC}{github.com/BaylorMultimediaLab/UI-LIC}.
\end{abstract}



\maketitle

\section{Introduction}

Learned image compression (LIC) has become an active research area, with many teams proposing new models, training procedures, entropy models, and perceptual optimization strategies. However, comparing LIC systems remains difficult in practice. Published results are often produced using different software stacks, training datasets, evaluation datasets, objective metrics, and implementation assumptions. As a result, differences in reported compression performance may reflect not only the underlying model, but also variations in training conditions, evaluation methodology, or experimental infrastructure.

This problem is particularly important for LIC because training data and optimization procedures are central to the final performance of a model. Unlike conventional image and video codecs, where a fixed codec implementation can often be evaluated directly on a shared test set, learned codecs require both training and evaluation to be considered part of the experimental protocol. 

To address this need, we present the \name, an open-source software framework for training, evaluating, and comparing learned image compression models. The framework is intended for LIC researchers and practitioners who need to compare models using common datasets, common evaluation scripts, and common reporting tools. It provides a unified interface for integrating model-specific training and evaluation code while preserving the flexibility needed to support heterogeneous LIC implementations.

\name{} currently integrates six learned image compression models through a common training and evaluation interface. Experiments are specified using JSON configuration files, which define the models, datasets, jobs, and relevant parameters. A dispatcher system parses these configurations and executes the selected training and evaluation tasks. The framework also provides objective metric reporting, including common quality metrics such as PSNR, SSIM~\cite{wang_image_2004}, VMAF~\cite{li2016vmaf}, and LPIPS~\cite{zhang2018lpips}, together with visual inspection tools for analyzing reconstruction quality and spatial error distributions. In addition to comparing LIC models with one another, the framework can be used to compare learned approaches against reference implementations of conventional codecs such as H.264/AVC~\cite{wiegand2003overview_h264_avc}, H.265/HEVC~\cite{sullivan_overview_2012}, and AV1~\cite{han_technical_2021}. This paper describes the motivation, design, main features, and intended use of \name~\cite{baylor2026licframework}.

\section{Background and Motivation}
Learned image compression (LIC) emerged from the idea of replacing hand-designed components of conventional image codecs with deep learning models trained for rate-distortion optimization. Instead of relying entirely on manually designed prediction, transform, and reconstruction tools, LIC models learn compact latent representations from data and reconstruct images using neural synthesis models. This data-driven approach has led to rapid progress, but it also introduces new challenges for reproducible evaluation.

Evaluating LIC models is more complex than evaluating conventional codecs. In traditional image and video compression, a fixed encoder and decoder can often be evaluated directly on a shared test set, and observed coding gains can usually be attributed to differences in codec design or encoder configuration. In LIC, however, the training dataset, training objective, model architecture, optimization procedure, checkpoint selection, and evaluation implementation can all influence the final rate-distortion performance. As a result, it is not always clear whether a reported gain comes from the compression model itself, from improved training data, from a different optimization objective, or from differences in the evaluation pipeline.

\begin{figure}
    \centering
    \includegraphics[width=0.75\linewidth,trim={0 4cm 0 9cm},clip]{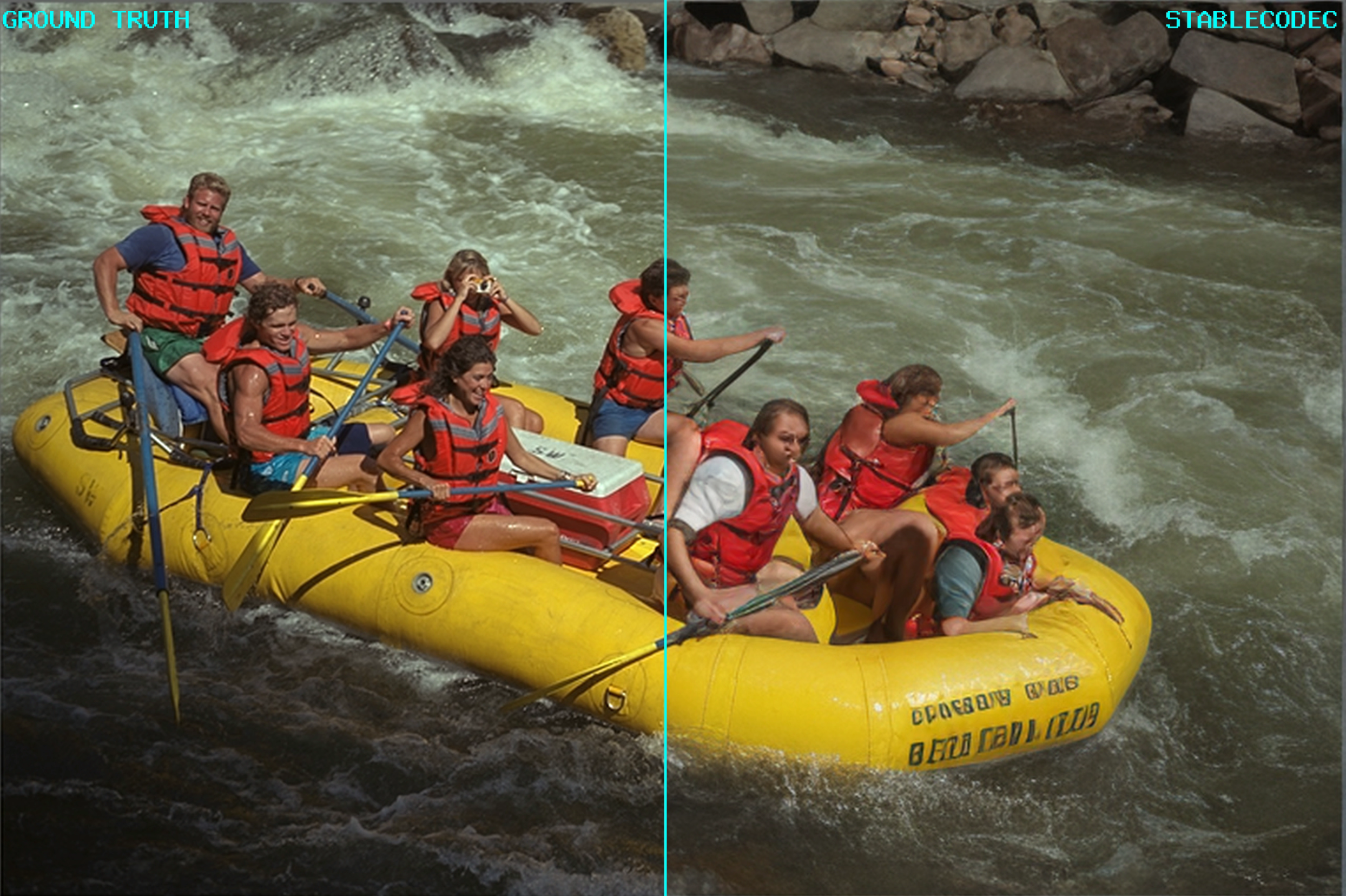}
    \caption{Before and after StableCodec's generative compression \cite{Zhang_2025_ICCV}. The resulting compressed image (right) has high visual quality and semantic similarity, but the faces are heavily distorted and the text is incoherent.}
    \label{fig:artifacts}
\end{figure}

Recent generative approaches further complicate this evaluation problem. Diffusion-based and other generative compression models may produce reconstructions that are visually plausible while differing from the reference image in semantically meaningful ways. These differences are sometimes described as hallucinations: content generated by the model that was not present in the original image. \cref{fig:artifacts} illustrates this type of artifact. Such examples highlight the need for evaluation tools that combine objective metrics with visual inspection, since high perceptual quality does not always imply faithful reconstruction of the source.

Objective quality metrics remain central to image and video codec evaluation because subjective human evaluation is costly, time-consuming, and difficult to reproduce. The most widely used distortion metric is peak signal-to-noise ratio (PSNR), which is simple to compute but does not account for important characteristics of human visual perception. More perceptually motivated metrics have therefore been proposed, including the structural similarity index measure (SSIM)\cite{wang_image_2004}, video multi-method assessment fusion (VMAF)\cite{li2016vmaf}, and learned perceptual image patch similarity (LPIPS)~\cite{zhang2018lpips}. However, different metrics can rank compression methods differently, especially when comparing models trained with different losses or perceptual objectives.

In practice, reproducing and comparing LIC models is also complicated by software heterogeneity. Existing models are often distributed as separate repositories with different dependencies, Python environments, configuration formats, training scripts, checkpoint conventions, preprocessing steps, and evaluation commands. Integrating several models into a single experimental study can therefore require substantial engineering effort before any scientific comparison can be performed.

These challenges motivate the Unified Interface for LIC, an open-source framework designed to train and evaluate multiple LIC models under common experimental conditions. The framework provides a unified interface for training models on shared datasets, computing common objective quality metrics, visually inspecting reconstructed images, and highlighting spatial artifacts identified by different metrics. The goal is to make LIC comparison more reproducible, while helping researchers analyze why different models perform well under some metrics but not others.

\subsection{Integrated Codecs}
To demonstrate the functionality of the framework, we include six Learned Image Compression models:
    Efficient Learned Image Compression (ELIC) \cite{jiang2022unofficialelic, he2022elic},
    StableCodec \cite{Zhang_2025_ICCV},
    TCM \cite{liu2023tcm},
    HPCM \cite{li2025hpcm},
    DCVC-RT intra \cite{jia2025towards}, and
    RwkvCompress \cite{Feng2025LALIC}.

Additionally, we include hooks for intra-frame coding with the following traditional codecs, via FFmpeg \cite{noauthor_ffmpeg_2024}:
    AVC/H.264 \cite{kalva_h264_2006} with \texttt{x264} and \texttt{NVENC};
    HEVC/H.265 \cite{sullivan_overview_2012} with \texttt{x265} and \texttt{NVENC}; and
    AV1 \cite{han_technical_2021} with \texttt{SVTAV1} and \texttt{NVENC}.


\section{Software Backbone}

Our framework is a modular, decoupled system that separates model training, inference, and visualization. We aim for simple usability, reproducibility, and extensibility. Our core software is implemented in Python 3.10.

\paragraph{Setup} With a single-command quick start script, the user can automatically set up a virtual environment for each model and download official pretrained weights. This also creates a virtual environment for the metrics evaluation, including a Docker container for the VMAF metric. Each model thus maintains the package versions specified by the original authors, so we do not break compatibility (e.g., with conflicting Python or CUDA versions).  

\paragraph{Argument Parsing Interface}
To simplify the execution of the various model training and evaluation scripts, we provide a base interface for argument parsing. [][] This interface provides a unified method for building training and testing commands, validating parameters, and executing the commands within various Conda environments set up for compatibility with implemented \ac{LIC} models.

\paragraph{Training and Testing Interfaces}

The training and testing interfaces for each LIC model inherit from a centralized base interface. With alias definitions in the base interface, we map model-specific variables to global arguments such as epoch count, learning rate, and dataset paths. Each interface enforces a list of additional required arguments that bypass default assignment, ensuring that mandatory parameters are supplied by the user prior to initialization.

\begin{figure*}
    \centering
    \includegraphics[width=0.95\linewidth]{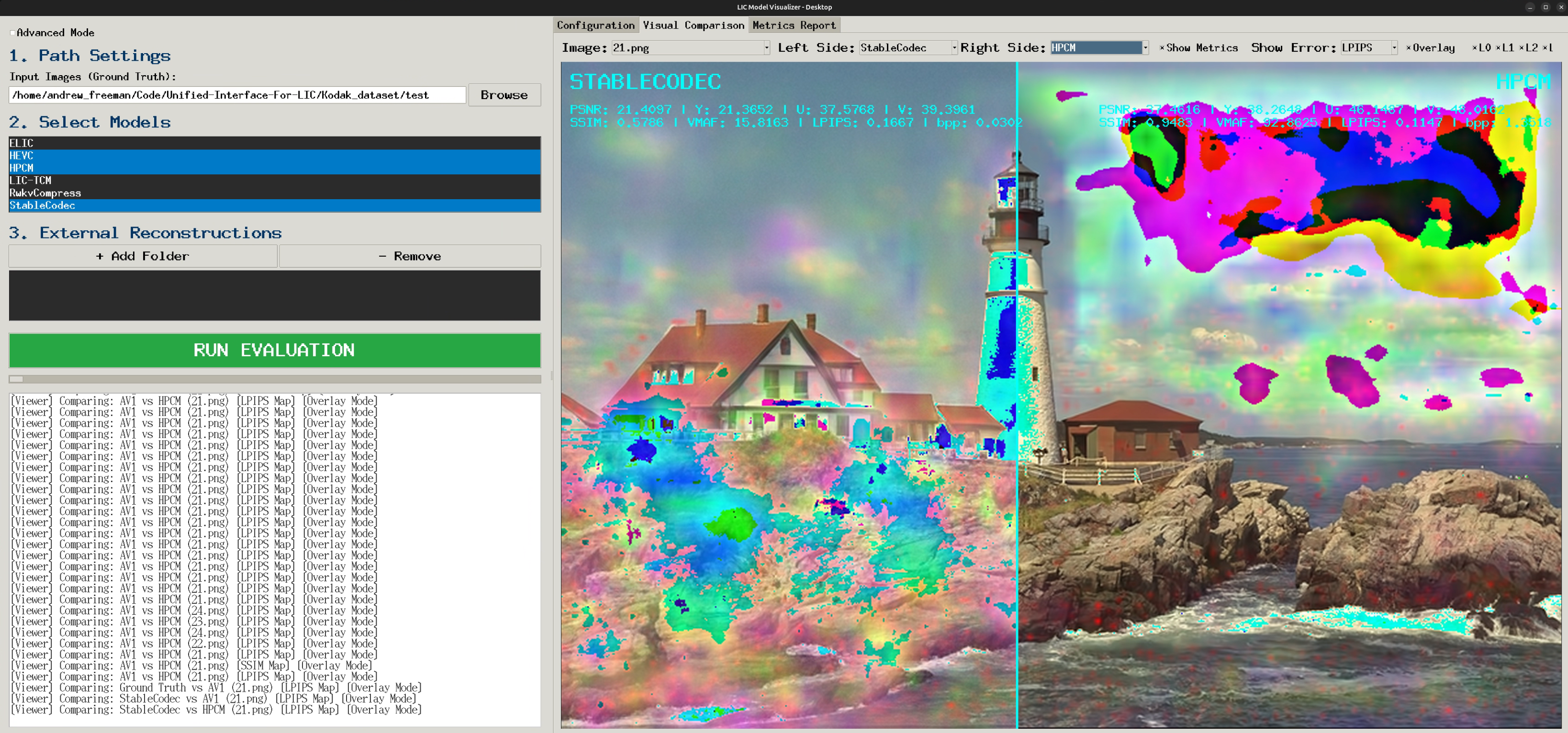}
    \caption{The GUI interface for our evaluation pipeline, showing LPIPS feature map overlays for two \ac{LIC}s. }
    \label{fig:gui}
\end{figure*}

\paragraph{Configuring Jobs}
Our job configuration script simplifies the construction of training and inference job argument files. The user can specify the models and parameters to be executed for each job. First, the user enters global arguments which are shared by all interfaces. Then, the user may optionally override our suggested default arguments for the remaining parameters of each model.

\paragraph{Job Dispatcher}
The dispatcher component  loads the interfaces from the provided interface directories along with an argument file that dictates the training and testing jobs to be executed. When an argument file has been loaded into the dispatcher, it check that all of the model's required arguments have been provided. It then prompts the user to choose which classes of jobs to execute. 

\paragraph{Model Implementation Changes}

Our primary modifications to the models' reference code are as follows:

\begin{itemize}
    \item \textbf{DCVC-RT intra:} Developed dedicated intra-frame training and inference scripts
    \item \textbf{StableCodec:} Standardized testing to use ImageFolder\\rather than H5Database. Implementation to convert epochs to steps so epochs can be passed to StableCodec's native training implementation.
    \item \textbf{ELIC and TCM:} Developed evaluation script to streamline saving bitstream and decoded representations.
    \item \textbf{All:} Replaced hard-coded file paths, added input validation, and extended error handling. 
\end{itemize}

In addition, we introduce evaluation driver scripts for each model. These scripts allowed us to unify the process of evaluating our image metrics by outputting encoded images into specified directories.

\section{Evaluation Pipeline}

Our evaluation pipeline spans \ac{LIC} model inference, traditional codecs, quality metric calculation, and error map generation. One can initiate an evaluation task in the command line using a JSON arguments file, or via our GUI program (\cref{sec:gui}).

\subsection{Metrics and Error Maps}\label{sec:error_maps}

The following metrics are integrated in each testing interface:
\begin{itemize}
    \item \textbf{Objective:} \ac{PSNR} (weighted and YUV components) and \ac{SSIM} \cite{wang_image_2004}
    \item \textbf{Perceptual:} \ac{VMAF} ~\cite{li2016vmaf} and \ac{LPIPS}~\cite{zhang2018lpips}
    \item \textbf{Efficiency:} Compressed bits-per-pixel (bpp), inference latency
\end{itemize}

In addition to an overall LPIPS score for each image, the evaluation script decomposes the five feature layers of the LPIPS AlexNet backbone. The lower layers roughly correspond to high-frequency features such as edges and fine textures, while the higher layers correspond to low frequency and semantic features. We save these feature maps as separate images to enable interactive error analysis in our GUI program (\cref{sec:gui}, below). We also produce error maps for the \ac{MSE}, block-based \ac{SSIM}, and normalized image gradients. 

\subsection{Rate-Driven QP Optimization for Traditional Codecs}\label{sec:qp_opt}

Some \ac{LIC} models do not inherently support variable-rate compression. In these cases, higher compression ratios may require additional sets of model weights. Traditional codecs, in contrast, provide granular rate-distortion control via a \ac{QP}. In our evaluation script, the user may enable \textit{bitrate equalization}. When enabled, the system first encodes the dataset with each \ac{LIC}. Then, for each image the smallest compressed bitrate is identified from the \ac{LIC} models' weights. This bitrate becomes the target for the traditional codecs. Each of these codecs may be invoked repeatedly in a \ac{QP} search process, until we have produced an encoded image as close as possible to the target bitrate. \cref{tab:kodak_quant_results} shows automated inference results from our framework on the Kodak dataset \cite{noauthor_true_2010} with bitrate equalization.Through the reported metrics, our software enhances observability for performance trade-offs between \ac{LIC} models and traditional codecs. For example, we can see that DCVC-RT offers the best all-around performance under the tested configurations, being the only model to outperform the traditional codecs on all quality metrics at an equal bitrate.

\subsection{Interactive Evaluation Tool}\label{sec:gui}

To streamline the usability of our framework and analysis tools, we provide a robust Graphical User Interface (GUI) as shown in \cref{fig:gui}. The GUI is implemented in Python with the Tkinter library.

\begin{table}[t]
\centering
\begin{tabular}{lccccc}
\hline
\textbf{Model} & \textbf{bpp} & \textbf{PSNR $\uparrow$} & \textbf{SSIM $\uparrow$} & \textbf{LPIPS $\downarrow$} & \textbf{VMAF $\uparrow$}\\
\hline
\underline{HPCM} & 1.25 & \textbf{42.23} & \textbf{0.952} & 0.07 & 92.58 \\
AV1 & 1.25 & 40.97 & 0.949 & 0.06 & \textbf{93.40} \\
AVC & 1.26 & 40.50 & 0.946 & 0.06 & 93.20 \\
HEVC & 1.25 & 40.83 & 0.949 & \textbf{0.06} & 92.63 \\ \hline
\underline{LIC-TCM} & 1.33 & \textbf{42.50} & \textbf{0.955} & 0.06 & 93.16 \\
AV1 & 1.33 & 41.29 & 0.953 & 0.06 & \textbf{93.79} \\
AVC & 1.33 & 40.83 & 0.949 & 0.06 & 93.59 \\
HEVC & 1.34 & 41.23 & 0.953 & \textbf{0.05} & 93.10 \\ \hline
\underline{RwkvComp.} & 1.27 & \textbf{42.29} & \textbf{0.953} & 0.06 & 93.01 \\
AV1 & 1.27 & 41.04 & 0.950 & 0.06 & \textbf{93.52} \\
AVC & 1.27 & 40.57 & 0.946 & 0.06 & 93.28 \\
HEVC & 1.28 & 40.97 & 0.950 & \textbf{0.05} & 92.80 \\\hline
\underline{StableCodec} & 0.03 & 27.56 & 0.572 & \textbf{0.17} & 22.62 \\
AV1 & 0.11 & 31.30 & 0.713 & 0.39 & 50.30 \\
AVC & 0.18 & \textbf{32.29} & \textbf{0.748} & 0.32 & \textbf{60.59} \\
HEVC & 0.13 & 32.14 & 0.741 & 0.33 & 56.15 \\ \hline
\underline{DCVC-RT} & 0.47 & \textbf{39.21} & \textbf{0.917} & \textbf{0.11} & \textbf{89.74} \\
AV1 & 0.47 & 36.23 & 0.877 & 0.16 & 82.56 \\
AVC & 0.47 & 35.72 & 0.866 & 0.16 & 82.00 \\
HEVC & 0.47 & 36.29 & 0.878 & 0.14 & 81.86 \\ \hline
\underline{ELIC} & 0.31 & \textbf{32.48} & 0.828 & 0.24 & \textbf{74.94} \\
AV1 & 0.31 & 32.15 & 0.832 & 0.22 & 74.80 \\
AVC & 0.31 & 31.76 & 0.819 & 0.22 & 74.25 \\
HEVC & 0.31 & 32.23 & \textbf{0.835} & \textbf{0.20} & 74.59 \\ \hline

\hline
\end{tabular}
\caption{Average codec performance on the Kodak dataset with pre-trained \ac{LIC} weights and bitrate equalization. }
\label{tab:kodak_quant_results}
\end{table}

\paragraph{Inference and Evaluation} To begin, the user selects a ground truth image dataset. Then, the user may selectively enable the desired \ac{LIC} models and traditional codecs to perform encoding. Optionally, the user may provide a path for a directory with existing image reconstructions; this allows one to derive the quality metrics and heat maps for codecs not yet integrated in our framework (e.g., codecs being developed by a researcher). By default, our interface exposes only the most common inference settings for each codec, such as potential \ac{QP} values and model weights paths. The user can enter ``advanced mode,'' however, which exposes additional settings. The program attempts to pre-fill the model weights path with the most likely option for each codec (e.g., if a weights filename includes the word ``best''). Once the user is satisfied with the configuration, a start button press initiates the inference commands for the selected codecs. If bitrate equalization is enabled (\cref{sec:qp_opt}), the traditional codecs are executed last.

\paragraph{Visualization} Once encoding, decoding, and metric evaluation are complete, the user can visualize the results. The user can select an image from the inference dataset, then select two codecs for a side-by-side comparison with a sliding mask. The left image defaults to the ground truth input. The metrics for a decoded image are displayed by default under the codec name. The user may choose to visualize an image error map (\cref{sec:error_maps}) produced during the metric evaluation stage. These can be viewed either as standalone images or as semi-transparent overlays atops the decoded images. Each layer for the LPIPS feature maps is represented with a distinct color, and each may be toggled independently. With the LPIPS error map visualizations enabled in \cref{fig:gui}, we can see in this example that StableCodec (left) performs better than HPCM (right) at representing the noise in the sky region, but performs worse at object representation due to its underlying generative model.

\paragraph{Metric Reports} Finally, we provide an interface to view the quantitative metric reports. Here, the user can see the performance of each codec across the entire dataset and compare the average results of the codecs.

\section{Conclusion and Future Work}
Learned image compression models are more difficult to evaluate than conventional codecs because training data, optimization objectives, model checkpoints, and evaluation pipelines can all influence the final compression performance. In addition, recent generative LIC models may introduce artifacts such as hallucinations, where the reconstructed image contains plausible visual content that was not present in the reference. These challenges motivate tools that support both reproducible objective evaluation and visual analysis.

In this paper, we presented the Unified Interface for LIC (\name), an open-source software framework for training, evaluating, and comparing learned image compression models under common experimental conditions. The framework enables multiple LIC models to be trained on shared datasets and evaluated using common objective quality metrics, while also providing visual tools to inspect reconstruction artifacts and spatial error patterns. It also supports comparisons with conventional image and video codecs, allowing learned models to be analyzed within a broader compression context.

The framework reduces the engineering effort required to reproduce and compare LIC models, which can save considerable time for researchers entering the field. By providing a common interface for training, evaluation, metric reporting, and visual inspection, \name{} aims to make LIC research more accessible, transparent, and reproducible.

Future work will focus on integrating additional LIC models, improving the visual analysis tools, and extending the framework to support more automated benchmark generation. We also plan to continue improving documentation and usability so that the framework can serve as practical shared infrastructure for the learned compression community.

\bibliographystyle{ACM-Reference-Format}
\bibliography{references}

\appendix

\end{document}